\documentstyle[11pt] {article}

\title{Gravity and Faster than Light Particles}

\author{Asher Yahalom$^{a,b}$ \\
$^a$ Isaac Newton Institute for Mathematical Sciences,\\
 20 Clarkson Road, Cambridge CB3 0EH, United Kingdom\\
$^b$ Ariel University, Ariel 40700, Israel\\
e-mails: asya@ariel.ac.il}

\begin{document}
\maketitle

\newcommand{\beq} {\begin{equation}}
\newcommand{\enq} {\end{equation}}
\newcommand{\ber} {\begin {eqnarray}}
\newcommand{\enr} {\end {eqnarray}}
\newcommand{\eq} {equation}
\newcommand{\eqs} {equations }
\newcommand{\mn}  {{\mu \nu}}
\newcommand{\sn}  {{\sigma \nu}}
\newcommand{\rhm}  {{\rho \mu}}
\newcommand{\sr}  {{\sigma \rho}}
\newcommand{\bh}  {{\bar h}}
\newcommand {\er}[1] {equation (\ref{#1}) }
\newcommand {\ern}[1] {equation (\ref{#1})}

\begin {abstract}

In this paper I discuss whether superluminal particles exist in
the general relativistic theory of gravity. It seems that the
answer to this question is negative. In truth the result may only
represent a difficulty to {\bf special}  but not general
relativity, the later allowing both Lorentzian
 and Euclidian metrics. An Euclidian metric does not restrict speed. Although only the Lorentzian metric is
stable \cite{Yahaloma}, an Euclidian metric can be created under special gravitational circumstances and persist in a
limited region of space-time causing possible superluminality.

\end {abstract}

\vfill

\section{Introduction}

It is well known that our daily space-time is approximately of
Lorentz (Minkowski) type with a metric $\eta_{\mn} = \ {\rm diag }
\ (1,-1,-1,-1)$. The above statement is taken as one of the
central assumptions of the theory of special relativity and has
been supported by numerous experiments. But one should ask why
should it be so?

Many textbooks \cite{Weinberg} state that in the general theory of relativity
any space-time is locally of the type $\eta_{\mn} = \ {\rm diag }
\ (1,-1,-1,-1)$, although it can not be presented so globally due
to the effect of matter. This is a part of the demands dictated by
the well known equivalence principle. The above principle is taken
to be one of the assumptions of general relativity other
assumption such as diffeomorphism invariance, and the requirement
that theory reduce to Newtonian gravity in the proper regime lead
to the Einstein equations:
\beq
G_\mn = -\frac{8 \pi G}{c^4} T_\mn
\label{ein}
\enq
in which $G_\mn$ is the Einstein tensor, $T_\mn$
is the stress-energy tensor, $G$ is the gravitational constant and $c$ is the velocity of light.

The Principle of Equivalence rests on the equality of gravitational and inertial mass,
demonstrated by Galileo, Huygens, Newton, Bessel, and E$\ddot{o}$tv$\ddot{o}$s. Einstein reflected that,
as a consequence, no external static homogeneous gravitational field could be detected
in a freely falling elevator, for the observers, their test bodies, and the elevator
itself would respond to the field with the same acceleration \cite{Weinberg}. This means that the observer will experience
himself as free, not feeling the effect of any force at all. Mathematically speaking for the observer
space time is locally (but not globally) flat and Minkowskian.

The point is that one need not assume that space-time is locally Lorentz
based on an empirical (unexplained) facts, rather one can derive this property from the field equations
based on the stability of the Minkowskian solution. Other unstable flat solutions of non Minkowskian type,
 such as an Euclidian metric $\eta_{\mn} = \ {\rm diag } \ (1,1,1,1)$ can exist in a limited region of space-time.
 In an Euclidian metric there are no speed limitations and thus the alleged particle can travel in faster than light speed.
The reader should notice that already Eddington \cite[page 25]{Edd} has considered the possibility that the
universe contains different domains in which some domains are locally
Lorentzian and others have some other local metric of the type
 $\eta_{\mn} = \ {\rm diag } \ (-1,-1,-1,-1)$ or the type
$\eta_{\mn} = \ {\rm diag } \ (+1,+1,\-1,-1)$. The stability of those domains was not discussed by Eddington.

Many authors have suggested explanations to the locally Lorentzian nature of space-time
 \cite{Greensite4,Greensite6,Itin,vanDam}. What is common to all the above approaches is that
 additional theoretical structures \& assumptions are needed.
In previous works \cite{Yahaloma,Yahalomb,Yahalomc} it was shown that General relativistic equations and
 linear stability analysis suffice to obtain a unique choice of the Lorentzian metric being the only one which is stable.
Other metrics are allowed but are unstable and thus can exist in only a limited region of space-time.
The analysis will not be repeated here, the reader is referred to the original literature.
It should be mentioned that the choice of coordinates in the Fisher approach to physics is also
justified using the stability approach \cite{Yahalomd}. The nonlinear stability question of the
Lorentzian metric was settled by D. Christodoulou \&  S. Klainerman \cite{Nonlin}.
As for the nonlinear instability of other spaces of constant metric this remains an open
question at this time.

The plan of this paper is as follows: in the first section we
describe possible mechanisms of metric change. In the following
section we describe a particle trajectory in a general flat space. The
next section includes analysis of particle trajectories in
Lorentz space-time for both the subluminal and superluminal cases.
The following section will discuss dynamics in the presence of an
Euclidean metric. Then the possible physical implications
of the current theory are described. Finally some concluding remarks are given.

\section{Possible Mechanisms of Metric Change}

It was shown in \cite{Yahaloma} that among the possible flat space metrics only
the Lorentzian metric is stable and can persist for a considerable region of space-time.
Nevertheless one may still inquire if a mechanism exists by which a metric change does occur\footnote{In the sense
that the eigen-values of the metric change signs.}, can we create some how a metric of the type $g_{\mn} = \ {\rm diag } \ (+1,+1,-1,-1)$
in some region of space-time? The answer obviously has to do with the only reason a metric should
 change according to \ern{ein} and this is $T_\mn$.
Looking at available solution of general relativity one finds that metric changes are quite common.

 The Schwarzschild square interval (in terms of spherical coordinates $t,r,\theta,\phi$) is given by:
\beq
c^2 d\tau^2 =
(1-\frac{r_s}{r})c^2 dt^2-\frac{dr^2}{1-\frac{r_s}{r}}-r^2(d\theta^2+\sin^2\theta
d\phi^2)
\label{Schwarzs}
\enq
In which $\tau$ is the proper time, and $r_s$ is the Schwarzschild radius (in meters) of the massive
body,
 which is related to its mass $M$ by $r_s = \frac{2GM}{c^2}$. It is obvious that while for $r>r_s$  the
 metric is locally (up to scaling) $g_{\mn} = \ {\rm diag } \ (+1,-1,-1,-1)$. For $r<r_s$  the
 metric is locally (up to scaling) $g_{\mn} = \ {\rm diag } \ (-1,+1,-1,-1)$. Hence the direction
 of temporal and (one) spatial axis is exchanged. Notice, however, that although the sign of the eigen-values
 did change we are still left with a Lorentzian metric.

 Another example is the Friedman-Lemaitre-Robertson-Walker square interval which is well known in cosmological models:
\beq c^2 d\tau^2 = c^2 dt^2-a(t)^2 \left( \frac{dr^2}{1-\kappa
r^2}+r^2(d\theta^2+\sin^2\theta d\phi^2)\right) \label{Friedman}
\enq $a(t)$ is known as the "scale factor" and $\kappa$ may be
taken to have units of length$^{-2}$, in which case $r$ has units
of length and a(t) is unitless. $\kappa$ is then the Gaussian
 curvature of the space at the time when $a(t) = 1$. Hence for radial distances such that $r<\frac{1}{\sqrt{\kappa}}$
 the  metric is locally (up to scaling) $g_{\mn} = \ {\rm diag } \ (+1,-1,-1,-1)$ that is Lorentzian. However, for
$r>\frac{1}{\sqrt{\kappa}}$ the  metric is locally (up to scaling)
$g_{\mn} = \ {\rm diag } \ (+1,+1,-1,-1)$. This means that a
particle propagating in a radial direction will experience an
Euclidean metric.

 One should notice that in the above cases a signature change is accompanied by a metric singularity \cite{Donald}
 while the signature changes considered by Eddington \cite{Edd} involve zeros. However, metric singularities are
 not curvature singularities and can be removed by proper choice of coordinates.

It will be also interesting to find a metric which is completely
Euclidean in some regime of space-time, while being Lorentzian in another
 such a transitory metric may take the form $g_\mn =  \ {\rm diag } \ (+1,2e^{-\frac{(x_\mu-x_{0\mu})^2}{\Delta^2}}-1,2e^{-\frac{(x_\mu-x_{0\mu})^2}{\Delta^2}}-1,
 2e^{-\frac{(x_\mu-x_{0\mu})^2}{\Delta^2}}-1)$ which is necessary to create an Euclidean domain of a width $\Delta$
 located at $x_{0\mu}$. More analytical effort should be invested in order to describe accurately the conditions under which space-time
 will become locally completely Euclidean.

\section{Particle Trajectories in Flat Space}

Let us now look at a particle travelling in a space-time with a constant metric of arbitrary form.
Such a particle can be described by the Action ${\cal A}$ and Lagrangian $L$:
\beq
{\cal A} = \int L d \tau, \qquad L= \frac{1}{2} m  u_\alpha u^\alpha + \frac{q}{c} u_\alpha A^\alpha
\label{Action}
\enq
In the above $\tau$ is some parameter along the trajectory, $x_\alpha$\footnote{Raising and lowering indices
is done using the metric as is customary.} are the particle coordinates,
$u_\alpha \equiv \frac{d x_\alpha}{d \tau}$, $m$ is the particle mass, $q$ is the particle charge and $A_\alpha$
are some functions of the particle coordinates (that transform as a four dimensional vector). Basic variational
analysis leads to the following equations of motion:
\beq
m \frac{d u^\alpha}{d \tau}= -\frac{q}{c} u^\beta (\partial_\beta A^\alpha - \partial^\alpha A_\beta)
\label{Eqmot}
\enq
It is customary to use as a parameter the length of the trajectory:
\beq
d \tau^2 = \left|\eta_{\alpha \beta} d x^\alpha d x^\beta \right|
\label{length}
\enq
in which $\eta_{\alpha \beta}$ is the metric.

\subsection{Lorentz Space-Time}

Let us assume a Lorentz Space-Time with a metric $\eta_\mn = \ {\rm diag } \ (1,-1,-1,\--1)$. Hence space-time
is dissected into spatial and temporal coordinates. The spatial coordinates are $\vec x = (x_1,x_2,x_3)$ and the
temporal coordinate is $x_0$. Since it is customary to measure time in different units (seconds) than space (meters)
we write $x_0 = c t$, in which $c$ serves as a units conversion factor. We now define the velocity: $\vec v \equiv \frac{d \vec x}{d t}$.
In a similar way we dissect $A_\alpha$ into temporal and spatial parts:
\beq
A_\alpha=(A_0,A_1,A_2,A_3) \equiv  (A_0, \vec A) \equiv (\frac{\phi}{c}, \vec A)
\label{Av}
\enq
Using \ern{Av}, we can define a magnetic field:
\beq
\vec B = \vec \nabla \times \vec A
\label{magnetic}
\enq
($\vec \nabla$ has the standard definition of vector analysis) and an electric field:
\beq
\vec E =-\frac{\partial \vec A}{\partial t} -\vec \nabla \phi
\label{electric}
\enq
For subluminal particles $v<c$ we can than write $d \tau^2$ as:
\beq
d \tau^2 = c^2 dt^2 (1-\frac{v^2}{c^2}), \qquad d \tau = c dt \sqrt{1-\frac{v^2}{c^2}}
\label{lengthlsl}
\enq
And using the above equations one can write the spatial part of \ern{Eqmot} as:
\beq
 \frac{d }{d t}\left(m \frac{\vec v}{\sqrt{1-\frac{v^2}{c^2}}}\right)= q \left(\vec E + \vec v \times \vec B\right)
\label{Eqmotsl}
\enq
The above equation shows clearly that a subluminal particle in a Lorentz space must remain subluminal. Since as the
particle is accelerated to $c$ its "effective mass" $m_{eff} \equiv  \frac{m}{\sqrt{1-\frac{v^2}{c^2}}}$ becomes infinite.
On the other hand for superluminal particles (which are $v>c$ at $\tau=0$) we can  write $d \tau^2$ as:
\beq
d \tau^2 = c^2 dt^2 (\frac{v^2}{c^2}-1), \qquad d \tau = c dt \sqrt{\frac{v^2}{c^2}-1}
\label{lengthlfl}
\enq
And using the above equations one can write the spatial part of \ern{Eqmot} as:
\beq
 \frac{d }{d t}\left(m \frac{\vec v}{\sqrt{\frac{v^2}{c^2}-1}}\right)= q \left(\vec E + \vec v \times \vec B\right)
\label{Eqmotfl}
\enq
Here the difficulty would be to go below the velocity $c$ (invalidating claims that the particle losses energy
 by interacting with the gauge field and becomes subluminal again).
In the absence of forces the velocity of the above
particle remains constant and superluminal. We conclude that in a Lorentz space time there is
a difficulty to pass the velocity $c$ from below or above as is well known.

\subsection{Euclidean Space-Time}

Let us assume an Euclidean space-time with a metric $\eta_\mn = \ {\rm diag } \ (+1,+1,\-+1,+1)$. Here space-time
is dissected (arbitrarily) into spatial and temporal coordinates as in the Lorentz space which are
measured in the customary units.  Again we define the velocity: $\vec v \equiv \frac{d \vec x}{d t}$ and dissect $A_\alpha$ into
temporal and spatial parts as in \ern{Av}.
Using \ern{Av}, we can define the magnetic field as in \ern{magnetic} but the electric field is defined now as:
\beq
\vec E =-\frac{\partial \vec A}{\partial t} +\vec \nabla \phi
\label{electrice}
\enq
notice that this definition for the electric field is different than in the Lorentz space but is necessary in order to
maintain Faraday's law.
For all particles either (subluminal or superluminal) we can than write $d \tau^2$ as:
\beq
d \tau^2 = c^2 dt^2 (1+\frac{v^2}{c^2}), \qquad d \tau = c dt \sqrt{1+\frac{v^2}{c^2}}
\label{lengthle}
\enq
And using the above equations one can write the spatial part of \ern{Eqmot} as:
\beq
 \frac{d }{d t}\left(m \frac{\vec v}{\sqrt{1+\frac{v^2}{c^2}}}\right)= q \left(\vec E - \vec v \times \vec B\right)
\label{Eqmote}
\enq
The above equation shows clearly that particles in an Euclidean space are quite indifferent to passing
the velocity $c$.

\section{Some Possible Physical Implications}

One obvious physical implication of the previous analysis is that a particle can be accelerated to a velocity close to the velocity
$c$ in a Lorentz space, enter into an Euclidean space and be accelerated further in this region to velocities
above the speed $c$ and emerge in a Lorentz space in which it will remain above the speed $c$ for ever
unless it is decelerated in an Euclidean space again.

This certainly may happen to a particle which travels radially in a Friedman-Lemaitre-Robertson-Walker metric
passing outwards the critical radius of $r_c=\frac{1}{\sqrt{\kappa}}$ and then coming back at superluminal velocities.

But if such particles do exist how would their existence bear on existing physical and astrophysical problems?

An obvious implication has to do with the homogeneity problem, superluminal particle are not restricted by
the velocity of light and hence can bring a very young universe into thermal equilibrium. Of course a more popular
mechanism for achieving this is inflation \cite{Inflation}. However, one should notice that a Higgs type fields do
not give the correct density perturbation spectrum \cite{Inflation}, hence one is forced to postulate a new field which
is not a part of any particle model and thus is a possible but inelegant solution of the homogeneity problem. Alternatively
one can speculate that homogeneity is achieved by ordinary matter which can become superluminal as the current analysis shows.

Another implication which is less obvious is that superluminal particle consist of at least some part of galactic or
inter-galactic dark matter \cite{DarkMatter1} (26.8\% of the matter in the universe are known to be dark).
Since a quantum theory of superluminal particles is not well developed at this stage, such a theory once elaborated may suggest that
those particles do not interact efficiently with radiation and thus appear dark.

A further implication has to do with the accelerating cosmological expansion. Since space-time has a different metric
for $r > r_c$ it may be that physics is different for such extreme distances. This bears on the correct interpretation of
red shifts in such extreme distances as well.

Last but not least one should remember that although classical physics is assumed to take place in a Lorentzian background,
quantum field theory calculations are done in an Euclidean background using the Wick rotation. This is usually justified on the basis
that it is  an analytic continuation. But an analytic continuation is a mathematical technique which has no physical justification
in Lorentzian space-time but makes perfect sense if part of space-time, in particular the part which is very close to the particle is Euclidean.
Hence one may speculate that each elementary particle may carry with it a "bubble" of a microscopic Euclidean space-time.

\section{Conclusion}

We have shown that general relativity allows for non-Lorentzian space-times in particular this is allowed
in part of the Friedman-Lemaitre-Robertson-Walker universe. The result of which is that superluminal particles can exist
in such a cosmology. Some of the cosmological implications of superluminal particles regarding the homogeneity problem,
and dark matter problems are underlined. Some other possible implications of non Lorentzian metrics which are not connected
to superluminality but may be a consequence of non-Euclidean metrics are also suggested. Of course much more detailed analysis
is needed to reach a definite conclusion regarding any of the above physical problems but the existence of non-Lorentzian space-times
and superluminal particles suggests a plausible solution.

\end{document}